# Emergent electric field control of phase transformation in oxide superlattices

Di Yi[1*], Yujia Wang[2], Olaf M. J. van 't Erve[3], Liubin Xu[4], Hongtao Yuan[5], Michael J. Veit[1,6], Purnima P. Balakrishnan[1,7], Yongseong Choi[8], Alpha T. N'Diaye[9], Padraic Shafer[9], Elke Arenholz[9,10], Alexander Grutter[11], Haixuan Xu[4*], Pu Yu[2,12,13*], Berend T. Jonker[3] and Yuri Suzuki[1,6]

1. Geballe Laboratory for Advanced Materials, Stanford University, Stanford, CA, 94305, USA
2. State Key Laboratory of Low Dimensional Quantum Physics and Department of Physics, Tsinghua University, Beijing, 100084, China
3. Materials Science and Technology Division, US Naval Research Laboratory, Washington, DC, 20375, USA
4. Department of Materials Science and Engineering, University of Tennessee, Knoxville, TN 37996, USA
5. National Laboratory of Solid-State Microstructures, College of Engineering and Applied Sciences, and Collaborative Innovation Center of Advanced Microstructures, Nanjing University, Nanjing 210093, China
6. Department of Applied Physics, Stanford University, Stanford, CA 94305, USA
7. Department of Physics, Stanford University, Stanford, CA 94305, USA
8. Advanced Photon Source, Argonne National Laboratory, Argonne, IL, 60439, USA
9. Advanced Light Source, Lawrence Berkeley National Laboratory, Berkeley, CA, 94720, USA
10. Cornell High Energy Synchrotron Source, Cornell University, Ithaca, NY 14853, USA
11. NIST Center for Neutron Research, National Institute of Standards and Technology, Gaithersburg, MD 20899-6102, USA
12. Frontier Science Center for Quantum Information, Beijing 100084, P. R. China
13. RIKEN Center for Emergent Matter Science (CEMS), Saitama 351-0198, Japan

* Emails: diyi.mse@gmail.com, hxu8@utk.edu, yupu@mail.tsinghua.edu.cn

**Abstract**

Electric fields can transform materials with respect to their structure and properties, enabling various applications ranging from batteries to spintronics. Recently electrolytic gating, which can generate large electric fields and voltage-driven ion transfer, has been identified as a powerful means to achieve electric-field-controlled phase transformations. The class of transition metal oxides (TMOs) provide many potential candidates that present a strong response under electrolytic gating. However, very few show a reversible structural transformation at room-temperature. Here, we report the realization of a digitally synthesized TMO that shows a reversible, electric-field-controlled transformation between distinct crystalline phases at room-temperature. In superlattices comprised of alternating one-unit-cell of $SrIrO_3$ and $La_{0.2}Sr_{0.8}MnO_3$, we find a reversible phase transformation with a 7% lattice change and dramatic modulation in chemical, electronic, magnetic and optical properties, mediated by the reversible transfer of oxygen and hydrogen ions. Strikingly, this phase transformation is absent in the constituent oxides, solid solutions and larger period superlattices. Our findings open up a new class of materials for voltage-controlled functionality.

**Introduction**

Electrolytes have been widely exploited to control electronic, magnetic and optical properties of materials when large electric fields are desired. Under electrolytic gating, changes in these properties have largely been attributed to changes in carrier concentration[1-5]. Recent advances have shown that electrolytic gating can also lead to the transfer of oxygen and hydrogen ions that modifies the structure of transition metal oxides (TMOs), leading to more dramatic modulation of their physical properties[6,7]. To fully exploit this mechanism, materials that can be electrically switched from one crystalline phase to another, with distinct physical properties, are highly desirable. Although ion transfer has been reported in many single-phase TMOs[8-18], reversible, electric-field-controlled transformation between distinct crystalline phases at room-temperature (RT) has been limited to a few systems, including binary oxides $VO_2$[8] and $WO_3$[12,13], and perovskite oxides $SrCoO_{3-\delta}$[14] and $SrFeO_{3-\delta}$[17] that exhibit topotactic transformations. Among these systems, coupled electronic, magnetic and optical phase transitions have only been demonstrated in $SrCoO_{3-\delta}$ thus far[14]. The scarcity of single-phase oxides as candidate materials demands new strategies to develop tunable materials to this end.

Here we show that digital oxide superlattices can be modulated by electrolytic gating to trigger reversible coupled phase transformations at RT. Using in-situ x-ray diffraction (XRD), we find that electrolytic gating induces a reversible structural transformation in superlattices comprised of alternating one-unit-cell of $La_{0.2}Sr_{0.8}MnO_3$ and $SrIrO_3$. By contrast, the constituent oxides, their solid solutions and larger period superlattices do not show such behavior. Through secondary ion mass spectrometry (SIMS), we confirm the transfer of both hydrogen and oxygen ions in the $[(La_{0.2}Sr_{0.8}MnO_3)_1(SrIrO_3)_1]_{20}$ superlattices. This transformation leads to a reversible metal-insulator transition accompanied by optical and magnetic transitions. More specifically, ferromagnetism with perpendicular magnetic anisotropy can be suppressed entirely in conjunction with metal-insulator and optical transitions. Together these results demonstrate the discovery of a new class of oxides that exhibits reversible and coupled phase transformations with rich functionalities through electrolyte-based ionic control.

## Results

**Synthesis of digital superlattices and reference films**

Perovskite TMOs of the $ABO_3$ form provide excellent candidates for strong electrochemical response due to the wide range of possible A and B cations in conjunction with the role of transition metal cation and oxygen anion (B-O) bonds for tunable electronic structure[19]. In addition, the correlated $d$ electrons of B cations give rise to strongly coupled physical properties. Here we studied epitaxial heterostructures of the 3$d$ TMOs La$_{1-x}$Sr$_x$MnO$_3$ that show rich physics and the 5$d$ TMO SrIrO$_3$, which has attracted recent research interests to explore exotic properties arising from spin-orbit coupling[20-26]. To elucidate the role of B-site cation ordering, we synthesized superlattices comprised of La$_{0.2}$Sr$_{0.8}$MnO$_3$ and SrIrO$_3$ ([(La$_{0.2}$Sr$_{0.8}$MnO$_3$)$_m$(SrIrO$_3$)$_m$]$_n$ with $m$=1, 2, 4), in-situ monitored by using reflection high-energy electron diffraction (RHEED) (Supplementary Fig. 1). In addition, we also synthesized single-phase films of La$_{0.2}$Sr$_{0.8}$MnO$_3$, La$_{0.7}$Sr$_{0.3}$MnO$_3$ and SrIrO$_3$ and B-site disordered solid solution films of Sr(Mn$_{0.5}$Ir$_{0.5}$)O$_3$ to shed light on the design strategies. Figure 1a shows the four types of materials. All samples are 40 unit-cells (approximately 16 nm) thick and were grown on (001)-oriented SrTiO$_3$ (STO) substrates. XRD shows that all samples exhibit good epitaxy and excellent crystallinity (Supplementary Fig. 2). Our previous studies have shown the excellent unit-cell layering of the superlattices by using transmission electron microscopy[22,27]. All samples were electrolytically gated with the ionic liquid DEME$^+$-TFSI$^-$ on the film surface, and voltages were applied across the electrolyte using electrical contacts to the top electrode and film surface (Figure 1a).

**Reversible structure transformation**

To directly observe structural changes induced by ionic liquid gating (ILG), we first performed in-situ XRD measurements during ILG at RT. First of all, upon ILG, in-situ XRD measurements show large changes in the out-of-plane lattice parameter ($\Delta c/c$) as high as 7% in [(La$_{0.2}$Sr$_{0.8}$MnO$_3$)$_1$(SrIrO$_3$)$_1$]$_{20}$ superlattices while single-phase films show much smaller changes and undergo irreversible structural changes at relatively small bias voltages. Secondly, only [(La$_{0.2}$Sr$_{0.8}$MnO$_3$)$_1$(SrIrO$_3$)$_1$]$_{20}$ superlattices show reversible structural transformations with the application of a reverse voltage. Figure 1b shows the time dependence of $\Delta c/c$ for different types of samples during voltage cycling; the (002) peak was monitored in-situ while incremental positive voltages were applied up to 2.5 V and then reversed to negative -3 V (Supplementary Fig. 2b-f). For manganate films, the maximum $\Delta c/c$ ratio is achieved at around 2% (1%) for $La_{0.2}Sr_{0.8}MnO_3$ ($La_{0.7}Sr_{0.3}MnO_3$) before the (002) peak disappears, and the process is irreversible even with the application of -3 V. For $SrIrO_3$ films, the (002) peak quickly decays at a lower voltage (~1.2 V). This may be attributed to the fact that the perovskite phase of $SrIrO_3$ is thermodynamically metastable in bulk and only epitaxially stabilized on STO substrates. Intriguingly, the maximum $\Delta c/c$ significantly increases to about 7% in solid solution films that have a random mixture of Mn and Ir cations. However, we find that the (002) peak in solid solution films also disappears below 2 V and that this change is irreversible.

By contrast, the B-site ordered $[(La_{0.2}Sr_{0.8}MnO_3)_1(SrIrO_3)_1]_{20}$ superlattices show a large reversible structural transformation. As shown in Figure 1b, the lattice expansion $\Delta c/c$ of superlattices is similar to that of solid solution films (~7%) up to bias voltages of 2 V. Beyond 2 V, the (002) peak of the superlattices retains the 7% expansion which can be reversed with negative voltages. Notably, as the period of superlattices ($m$) increases, the lattice expansion decreases and the change becomes irreversible (Supplementary Fig. 3), further highlighting the importance of B-site cation ordering at the atomic scale. For the rest of this paper, the term superlattice is used to refer to $[(La_{0.2}Sr_{0.8}MnO_3)_1(SrIrO_3)_1]_{20}$ superlattices unless otherwise stated.

Figure 1c shows the in-situ XRD results of the superlattices during repeated cycling. With +2.5 V, the (002) peak (around 45.7°, corresponding to $c \sim 0.396$ nm) gradually shifts towards lower angles, indicating the elongation of the $c$ lattice parameter. Eventually, a new diffraction peak develops at around 42.4° (corresponding to $c \sim 0.425$ nm), revealing the emergence of a new phase (denoted as phase B). By reversing the voltage to -3 V, the superlattice gradually returns to the original phase (denoted as phase A). Notably, the ILG-induced changes are nonvolatile, as phase B is stable over an extended time at ambient conditions (Supplementary Fig. 4a). Thus, we can characterize the two phases by ex-situ probes. Figure 1d shows full-range XRD of the superlattices, showing good epitaxy and cation ordering (deduced from pronounced satellite peaks) in both phases. Reciprocal space mapping shows that both phases are coherently strained by the underlying substrate (Supplementary Fig. 5).

**Voltage-driven dual-ion transfer**

To identify whether the transfer of ions plays a role, we employed SIMS to measure ex-situ the depth profiles of oxygen and hydrogen. It is known that ILG can induce changes in both oxygen and hydrogen stoichiometries in oxides, and one possible origin is the electrolysis of residual water[10,14]. To probe hydrogen ions, we stabilized two superlattices with phases A and B by ILG. As shown in Figure 2a, SIMS results of the superlattice in phase B (red) reveal a significantly higher concentration of hydrogen than that of phase A (blue), confirming the incorporation of hydrogen ions in phase B. To explore the possibility of oxygen ion extraction (oxygen vacancy creation), we first stabilized two superlattices with phases A and B by ILG. Then both samples were thermally annealed in $^{18}O_2$ gas (Supplementary Fig. 4b). We find that the phase B sample after thermal annealing in $^{18}O_2$ shows a dramatic increase in $^{18}O$ signal throughout the entire sample, while the phase A sample only shows an increase near the surface (Figure 2b). Although the depth profile of the $^{18}O$ signal does not directly measure the amount of oxygen vacancies, the additional incorporation of $^{18}O$ throughout the entire phase B sample indicates that ILG creates oxygen vacancies in phase B (Supplementary Fig. 4c). It is noted that the SIMS results are consistent with polarized neutron reflectivity results, revealing the accumulation of oxygen vacancies and hydrogen ions in phase B (Supplementary Fig. 6). Therefore, we demonstrate that the phase transformation is mediated by voltage-driven dual-ion transfer, characterized by the exchange of both oxygen and hydrogen ions[14].

**Valence modulation**

The large structural changes induced by ILG should be accompanied by changes in cation valence and oxygen bonding. To probe valence changes, we carried out ex-situ X-ray absorption spectroscopy (XAS) at the Mn L-edge, Ir L-edge and O K-edge. The absorption peak at the Mn-L edge (Figure 3a and Supplementary Fig. 7a-c) shifts to lower energy by ~2.1 eV when comparing phase B to phase A, indicating a substantial decrease in Mn valence[28]. By comparing peak positions and multiplet features to references (Supplementary Fig. 8), we find that the Mn cations are close to +3.5 oxidation state in phase A and are dominated by +2 oxidation state in phase B. It is noted that the change of Mn valence in the superlattices is much larger than that in $La_{0.2}Sr_{0.8}MnO_3$ films under ILG (Supplementary Fig. 9), consistent with the magnitude of the structural change. Complementary results at the Ir-$L_3$ edge (Figure 3b and Supplementary Fig. 7d) also show a clear shift towards lower energy in phase B, revealing a large decrease of the Ir oxidation state[29]. Moreover, the oxygen K-edge results (Figure 3c) exhibit a complete suppression of the spectral feature associated with the hybridization of oxygen and Ir/Mn cations in phase B, confirming impact on the valence state. We also note that distinct XAS features around 540 eV were observed between two phases of the superlattices, which are known to indicate the presence of hydroxyl groups and are consistent with the SIMS results[30].

**Voltage control of physical properties**

Strikingly the large structural and chemical changes are accompanied by a concurrent metal-insulator transition, suppression of ferromagnetism and enhancement of optical transparency. We first studied the transport properties under ILG (Figure 4a). The resistivity changes between the two phases are reversible during repeated cycling at RT (Figure 4b). As shown in Figure 4c, the superlattice in phase A shows the lowest resistivity and a weak temperature dependence. With increasing positive voltage, the resistivity gradually increases and eventually saturates in a highly insulating state of phase B (Supplementary Fig. 10). The resistivity changes by about two orders of magnitude at RT and three orders of magnitude around 200 K. The large tuning range and high reversibility of resistivity, mediated by voltage-driven ion transfer, make the superlattices of potential interest in resistive random-access memory devices[31], electrochemical sensors[15] and neuromorphic computing[32].

The metal-insulator transition is accompanied by a significant modulation of optical transparency at visible wavelengths under ILG, showing an electrochromic effect in the superlattices. Figure 4d shows ex-situ optical transmittance measurements in the visible and near-infrared regions. These measurements exhibit a reversible modulation of optical transmittance around 25-30 % over the entire wavelength (400 to 1800 nm) between the two phases. This modulation can also be directly observed by eye, noting the strong opacity changes between the two phases (Figure 4d inset). This electrochromic effect may find applications in smart windows[33].

The most intriguing effect due to ILG was found in the suppression of ferromagnetism and perpendicular magnetic anisotropy (PMA). We note that the PMA, which is challenging to realize, is an emergent interfacial phenomenon in the superlattices[27]. The change of magnetism was confirmed by both in-situ and ex-situ magnetic characterization. The in-situ measurements exploited the magneto-optic Kerr effect (Figure 5a and Supplementary Fig. 11) and showed that ILG can fully suppress ferromagnetism with strong PMA. To quantify the changes over a wider temperature range, we carried out ex-situ measurements by using SQUID magnetometry (Figure 5b and 5c). When the superlattice is in phase A, a ferromagnetic ground state is stabilized with saturation magnetization $M_s \sim 2$ $\mu_B$/Mn, magnetic easy axis along the out-of-plane direction and Curie temperature ($T_c$) around 150K. With increasing positive voltage, both $M_s$ and $T_c$ decrease until the ferromagnetic transition is fully suppressed when the superlattice is in phase B. The suppression of ferromagnetism is also highly reversible after repeated cycling (Supplementary Fig. 12). Voltage-controlled PMA has been shown to be critical in developing magnetic memory devices with high density, good stability and low power-consumption[34,35].

## Discussion

By combining multiple in-situ and ex-situ probes, we have shown that the digital superlattices with atomic layering structure provide good candidates for electrolyte-based ionic control with both wide tunability and good reversibility. The structural and chemical characterization clearly show a reversible, electric-field-controlled transformation between two crystalline phases of the superlattices at RT, with distinct lattice parameters, chemical stoichiometries and valences, as schematically shown in Figure 1e. More specifically, phase A of the superlattices is close to the as-grown state in terms of the lattice parameter ($c \sim 0.396$ nm) and chemical stoichiometry (close to $ABO_3$). On the other hand, phase B shows a largely expanded lattice ($c \sim 0.425$ nm) and a massive loss (gain) of oxygen (hydrogen) ions ($H_xABO_{3-\delta}$). Although it is challenging to precisely quantify the local concentration of ions, we note that the magnitude of lattice expansion (~7%) and valence change are comparable to the highest reported values in singe-phase perovskite TMOs (e.g. $H_xSrCoO_{3-\delta}$ with large transfer of hydrogen ($x \sim 1$) and oxygen ($\delta \sim 0.5$) ions)[14]. This phase transformation is accompanied by simultaneous modulation of the electronic, optical and magnetic properties.

Our results have revealed insights into the strategies to develop complex oxides for electrolyte-based ionic control. Firstly, as compared to single-phase constituent oxides, both superlattices and solid solution films exhibit giant structural changes with the application of positive voltages. This result indicates that the intermixing of $3d$ and $5d$ transition metal cations can modify the enthalpy of formation energy and diffusivity of ions (oxygen vacancies and hydrogen ions)[36]. More importantly, in order to achieve wide tunability and high reversibility, candidate materials need to be transformed into other crystalline phases after massive ion transfer, rather than undergoing phase decomposition by electrolytic gating. Our results have shown that a new crystalline phase only appears in the $[(La_{0.2}Sr_{0.8}MnO_3)_1(SrIrO_3)_1]_{20}$ superlattices after ion transfer and not in the parent compounds, solid solutions or larger period superlattices, thereby highlighting the critical role of B-site cation ordering at the atomic scale.

Similarities can be drawn between these atomically layered $(La_{0.2}Sr_{0.8}MnO_3)_1(SrIrO_3)_1$ superlattices and single-phase oxides $SrCoO_{3-\delta}$[14] and $SrFeO_{3-\delta}$,[17] where structural phase transformations are brought about by electrolytic gating. Both $SrCoO_{3-\delta}$ and $SrFeO_{3-\delta}$ show a topotactic transformation into a vacancy-ordered brownmillerite phase after the transfer of ions, thus preserving the lattice framework without losing the crystallographic orientation and lattice structure[14,17,37]. Although single-phase manganate and iridate do not show this transformation at room-temperature as shown in Figure 1b and Supplementary Fig. 2, the digital superlattices can provide a lattice framework to facilitate a structural phase transition. This is closely related to the unit-cell layered structure of the superlattices, which leads to the different formation energies of ions depending on the local chemical environment.

To reveal the correlation between the formation energy of ions and local chemical environment, we performed first-principles calculations. For simplicity, we use the supercell that is composed of alternating one-unit-cell of $SrIrO_3$ and $SrMnO_3$ in the calculations (details are included in the Methods section and Supplementary Note 13). Figure 6a shows the calculated crystal structure of the superlattice without ionic defects, showing different rotation angles of $IrO_6$ and $MnO_6$ octahedra along the **c** axis. Subsequently, an oxygen vacancy or a hydrogen interstitial is introduced into the supercell and different possible positions have been considered. For instance, given the unit-cell layered structure, three types of oxygen vacancy positions were tested, i.e., in the SrO layer, in the $IrO_2$ layer, or in the $MnO_2$ layer. These sites are labeled as O1, O2, and O3 in Figure 6b. Our calculation results reveal that these sites indeed show different vacancy formation energies and suggest that an oxygen vacancy is energetically more favorable in the $MnO_2$ layer (O3) (Supplementary Table 1). Further calculations reveal that the formation energy of hydrogen interstitials also depends on the local chemical environment (Supplementary Fig. 13 and Note 13). Therefore, the digital superlattice is highly likely to develop certain types of ion ordering after the transfer of ions, leading to a reversible transformation into another crystalline phase instead of phase decomposition.

Our results suggest that the digital superlattices provide a new class of materials for electrolytic control of structure and function. The atomically layered structure is found to be critical to enable reversible phase transformation. The underlying mechanism may be correlated to the strength of cation-oxygen bonds arising from the 3$d$ versus 5$d$ cations as well as the different rotation/distortion of oxygen octahedra. These oxygen octahedral rotations have been predicted by our first principles calculations to exhibit different magnitudes across the interfaces (Figure 6a). This digital synthesis approach can be applied to the family of TMOs, which includes but is not limited to iridate and other 3$d$ TMOs. Further studies on additional material systems would provide a more comprehensive understanding of the factors governing these structural transformations. These structural transformations can also affect physical properties of the digital superlattices, including emergent phenomena appearing at these oxide interfaces[38]. Therefore, the digital synthesis approach also offers a means to tune emergent functionalities at interfaces through electrolyte-based ion transfer.

In conclusion, we have shown a reversible, electric-field-controlled transformation between distinct phases at RT in atomically layered oxide superlattices, with coincident electronic, optical and magnetic phase transitions. By contrast, these phenomena are not otherwise observed in the individual constituent oxides, solid solutions or larger period superlattices. Our findings reveal a hitherto-unexplored strategy to develop complex oxides, in which the electrolyte-based ionic control can provide extensive tunability that can be harnessed in electronic/spintronic[6,39], energy and environmental applications[19].

**Methods**

**Growth of oxide heterostructures**

Thin films and superlattices were grown on atomically flat SrTiO$_3$ (STO) (001) substrates by pulsed laser deposition using a KrF excimer laser operating at 248 nm and multiple targets with corresponding chemical stoichiometry. The laser fluence was 0.9 J cm$^{-2}$ and the repetition rate was 1 Hz. The growth temperature and oxygen partial pressure were maintained at 700 °C and 50 mTorr (6.7 Pa). Before growth, STO substrates were prepared in a buffered hydrofluoric acid etch and subsequently annealed at 1,000 °C for 3 hours to generate atomically flat surfaces as confirmed by atomic force microscopy. During growth, the samples were in-situ monitored by using reflection high-energy electron diffraction (RHEED), showing layer-by-layer growth for both La$_{0.2}$Sr$_{0.8}$MnO$_3$ and SrIrO$_3$. For manganate films and superlattices, RHEED oscillations were observed over the entire deposition process, leading to precise control of interface quality and overall thickness. For iridate and solid solution films, RHEED patterns decay after 5-10 oscillations and the overall thickness was controlled by counting the total number of laser pulses. After growth, the samples were cooled down to room temperature at a rate of ~10 °C per minute in 50 mTorr (6.7 Pa) oxygen.

**In-situ and ex-situ XRD measurements**

XRD measurements were performed by using a high-resolution diffractometer using monochromatic Cu K$_{\alpha 1}$ radiation ($\lambda$ = 1.5406 Å). Before ionic liquid gating (ILG), the edges of the samples (5×5 mm$^2$) were covered with a gold electrode by dc sputtering to form the bottom electrode. Subsequently, conductive gold or platinum wires were connected to the electrode by painting conductive silver adhesive. A thin platinum plate was used as the top electrode. During in-situ XRD measurements, the samples were first aligned with the substrate (002) peak without the ionic liquid; then, a small drop of ionic liquid (commercial ionic liquid DEME$^+$-TFSI$^-$) was added to cover both the film surface and the platinum plate. The voltage between the bottom electrode and the top electrode was initially set to 0 V and then ramped to desired values. Meanwhile, the XRD spectra were collected continuously with a scanning rate of 3° per minute. For ex-situ XRD measurements, the samples were gated by ILG in the same configuration. After being transformed into the desired phases, the samples were rinsed several times with acetone and isopropanol to remove the ionic liquid residue before performing XRD scans (such as Figure 1d). It is noted that the same cleaning procedure was applied for other ex-situ measurements.

**Ex-situ SIMS measurements**

To directly determine the changes in hydrogen and oxygen stoichiometry, we carried out secondary-ion mass spectrometry (SIMS) measurements (using an instrument from IONTOF GmbH). The mass resolution is about 4,000 atomic mass units (full-width at half-maximum). During the measurement, the Cesium-ion beam (2 KeV) was rastered over a region of about $250 \times 250$ μm$^2$ but data were collected only in an area of $50 \times 50$ μm$^2$ within that region to avoid disturbance from the crater edge.

**Ex-situ polarized neutron reflectivity measurements**

To characterize the structural and magnetic depth profile, we performed polarized neutron reflectometry (PNR) on as-grown and gated superlattices by ILG in which 2.5V was applied for 20 minutes. We measured the spin-dependent reflectivity as a function of momentum transfer $Q$ along the film normal after field cooling to 30 K in an applied magnetic field of 3 T. Because the spin-dependent reflectivity is a function of both the in-plane magnetization and nuclear composition, both the structural and magnetic depth profiles of the superlattices can be reconstructed by fitting the data using the Refl1D software program for $\chi^2$ optimization[40]. Due to the strong perpendicular magnetic anisotropy of the superlattices, it is expected that any component of the net magnetization not aligned along the applied field will instead orient along the film normal, where the neutron scattering selection rules render it invisible. Thus, no in-plane magnetization component is expected to be perpendicular to the applied field, and no spin-flip scattering is expected. We therefore collected only the non-spin-flip reflectivity spectra ↑↑ and ↓↓, where the arrows represent neutrons parallel and antiparallel to the applied field respectively.

**Ex-situ X-ray absorption spectroscopy**

We carried out X-ray absorption spectroscopy (XAS) measurements at the Mn L-edge and O K-edge at beamlines 4.0.2 and 6.3.1 of the Advanced Light Source at Lawrence Berkeley National Lab. The angle of the incident x-rays is 30 degrees to the sample surface. To probe Mn in the superlattices, we employed two detection modes: total-electron-yield (TEY) mode that is sensitive to the surface and luminescence-yield (LY) mode that probes the entire sample (Supplementary Fig. 7b and 7c). The two detection modes show similar x-ray absorption spectra, suggesting that the valence changes occur over the entire sample. For the oxygen K-edge, we only employed the TEY mode to avoid the large contributions from the oxide substrates. XAS at the Ir $L_3$-edge was measured at beamline 4-ID-D of the Advanced Photon Source at Argonne National Lab. The results were taken by collecting the fluorescence-yield (FY) signal with a grazing incidence geometry.

**In-situ transport measurements**

Transport measurements were performed in a Quantum Design Dynacool system using a four-probe electrical contact geometry shown in Figure 4a. The samples were cut into a rectangular shape ($2\times5$ mm$^2$). Patterned electrodes (AuPd, ~50 nm) were deposited on the surface of the superlattice by dc sputtering. Then a thin Pt foil was used as the top electrode. Bias voltage ($V_g$) was applied between the top electrode and the superlattice, contacted by a drop of ionic liquid (DEME$^+$-TFSI$^-$). The source-drain current was set to be 2 μA for all measurements. To measure magnetoresistance, we applied magnetic fields of up to 7 T in the out-of-plane direction. The transport measurements were performed under vacuum in the cryostat (Dynacool system chamber).

**Ex-situ optical transmittance measurements**

We used superlattices grown on double-side polished STO (001) substrates (10×10 mm$^2$) for optical transmittance measurements. A bare double-side polished STO (001) substrate was processed through the same thermal cycling as the other samples and was used as a reference for the optical transmittance measurements. Ex-situ optical transmittance spectra were taken in air at room-temperature with spectrophotometers (Agilent Cary 6000i UV/Vis/NIR) which cover the visible and near infrared range with wavelengths between 400 nm and 1800 nm.

**In-situ and ex-situ magnetic measurements**

In-situ magneto-optic Kerr effect (MOKE) measurements were performed by using a lateral top gate electrode as shown in Supplementary Fig. 11a. The gate electrode was deposited on an insulating LaAlO$_3$ (LAO) single crystal, which was mounted next to the superlattice on an insulating polymer platform. The measurements were carried out in vacuum in a cryostat. Liquid nitrogen was used to cool down the superlattice.

To obtain quantitative results over a wider temperature range, a Quantum Design 7T-SQUID magnetometer was employed to perform ex-situ magnetic measurements. The magnetic hysteresis loops were measured at 10 K after field cooling in 7 T. The temperature dependent magnetization was measured during the warming process with an applied field of 0.2 T. Magnetic properties were measured both along the out-of-plane (Figure 5) and in-plane ([100], Supplementary Fig. 12) directions.

**First-principles calculations**

We performed non-collinear spin-resolved density-functional-theory (DFT) calculations using the Vienna Ab initio Simulation Package (VASP)[41,42] to estimate the defect formation energy of an oxygen vacancy or a hydrogen interstitial. The PBEsol functional[43] was used in the form of the projector augmented wave method[44]. DFT+U approach developed by Dudarev et al[45] was adopted to describe the correlation effects of the system. Additional information about our DFT calculations is included in the Supplementary Note 13.

For simplicity, we neglected the small amount of La doping in the superlattices and used the $SrIrO_3/SrMnO_3$ (SIO/SMO) supercell in the calculation, which still reveals the role of B-site cation ordering. The supercell is composed of alternating one-unit-cell of SIO and SMO stacking in the out-of-plane direction while the in-plane direction contained double perovskite cells to account for octahedral rotation, resulting in a total chemical formula of $Sr_4Ir_2Mn_2O_{12}$. Subsequently an oxygen vacancy or a hydrogen interstitial was introduced into the superlattice (see Supplementary Note 13 for more details). All structures were relaxed until the electronic convergence of $10^{-6}$ eV was reached and the force on each atom was smaller than 0.01 eV Å$^{-1}$. The energy cut-off of the plane-wave basis was set as 550 eV, and the 6×6×4 Gamma-centered Monkhorst-Pack grid was employed in the calculations.

## Data Availability

The data that support the findings of this study are available from the corresponding authors upon reasonable request.

## Acknowledgements

The authors acknowledge experimental support from B. Kirby and J. Liu. The work at Stanford by D.Y. was supported by the Air Force Office of Scientific Research (AFOSR) under Grant FA9550-16-1-0235. M.J.V. was supported by an NSF Graduate Fellowship and Vannevar Bush Faculty Fellowship program sponsored by the Basic Research Office of the Assistant Secretary of Defense for Research and Engineering and funded by the Office of Naval Research through grant N00014-15-1-0045. P.P.B. was supported by the National Science Foundation under Grant No.1762971. Work at Tsinghua was financially supported by the Basic Science Center Project of NFSC under grant No. 51788104; the National Basic Research Program of China (grants 2015CB921700 and 2016YFA0301004); the Beijing Advanced Innovation Center for Future Chip (ICFC). Work at NRL was supported by core programs at the Naval Research Laboratory. L.X and H.X was supported by the Organized Research Unit Program (ORU-IMHM-18) at the University of Tennessee (UT) and utilized computing resources at the Advanced Computing Facilities of UT. This research used resources of the Advanced Light Source, which is a DOE Office of Science User Facility under contract no. DE-AC02-05CH11231. This research used resources of the Advanced Photon Source, a U.S. Department of Energy (DOE) Office of Science User Facility operated for the DOE Office of Science by Argonne National Laboratory under Contract No. DE-AC02-06CH11357. Part of this work was performed at the Stanford Nano Shared Facilities (SNSF), supported by the National Science Foundation under award ECCS-1542152. Certain trade names and company products are identified to specify adequately the experimental procedure. In no case does such identification imply recommendation or endorsement by the National Institute of Standards and Technology, nor does it imply that the products are necessarily the best for the purpose.

## Author Contributions

D.Y. and Y.S. conceived the project and designed the experiments. D.Y. fabricated the samples. D.Y. performed the ex-situ X-ray diffraction, magnetic and optical measurements with M.J.V. and P.P.B. Y.W. and P.Y. performed the in-situ X-ray diffraction and secondary ion mass spectroscopy measurements. O.M.J. and B.T.J. performed the in-situ magnetic optical Kerr effect measurements. D.Y. and H.Y. performed the electrical transport measurements. L.X. and H.X. performed the first-principles calculations. D.Y. performed the soft X-ray absorption measurements with the support of A.T.N., P.S. and E.A., and hard X-ray absorption measurements with the support of Y.C. A.G. performed the polarized neutron reflectivity measurements. D.Y. and Y.S. wrote the manuscript and all authors commented on it.

## Competing Interests

The authors declare no competing interests.

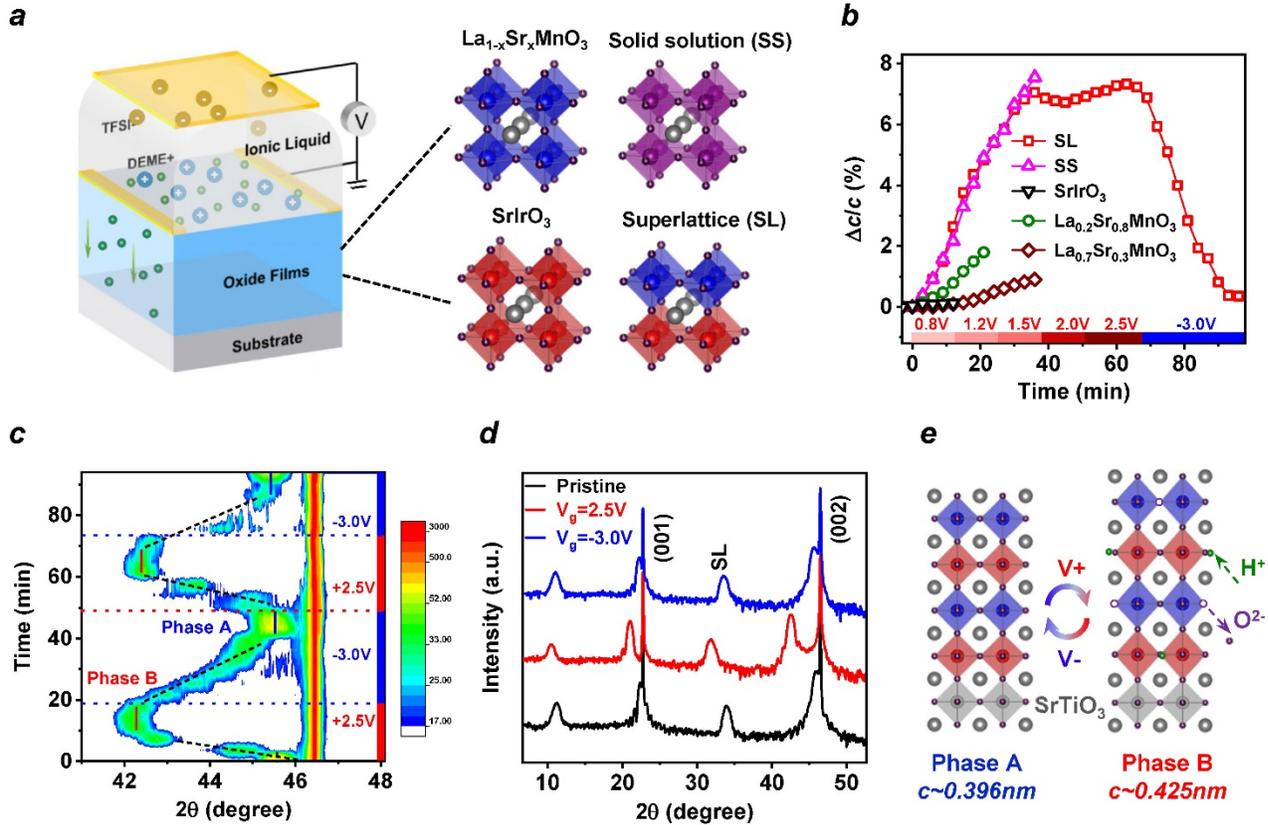

**Figure 1 | Electric field control of structural transition.** (a) Schematic of ionic liquid gating (ILG) that induces the ion transfer between oxides and ionic liquid. Four types of oxides were studied, i.e. manganate ($La_{1-x}Sr_xMnO_3$) and iridate ($SrIrO_3$) films, solid solution ($Sr(Mn_{0.5}Ir_{0.5})O_3$) films and superlattices ($[(La_{0.2}Sr_{0.8}MnO_3)_m(SrIrO_3)_m]_n$) on $SrTiO_3$ substrates. (b) Modulation of out-of-plane lattice constant ($\Delta c/c$) during voltage cycling (with incremental positive voltages to +2.5 V and -3.0 V) for different samples (SS refers to $Sr(Mn_{0.5}Ir_{0.5})O_3$ and SL refers to $[(La_{0.2}Sr_{0.8}MnO_3)_1(SrIrO_3)_1]_{20}$). The lattice constant was extracted from in-situ measurements of the (002) peak position. The end of the curves for all samples except the superlattice corresponds to phase decomposition above certain voltages. (c) In-situ X-ray diffraction results of the superlattices around the (002) peak during ILG in repeated voltage cycling (sequence of +2.5 V, -3.0 V, +2.5 V, -3.0 V), showing a reversible electric-field-controlled transformation between two phases of the superlattices. (d) Full-range X-ray diffraction of a superlattice in as-grown pristine state, positively gated state and reversibly gated state. (e) Schematic of the reversible phase transformation mediated by dual-ion (both $H^+$ and $O^{2-}$) transfer.
22

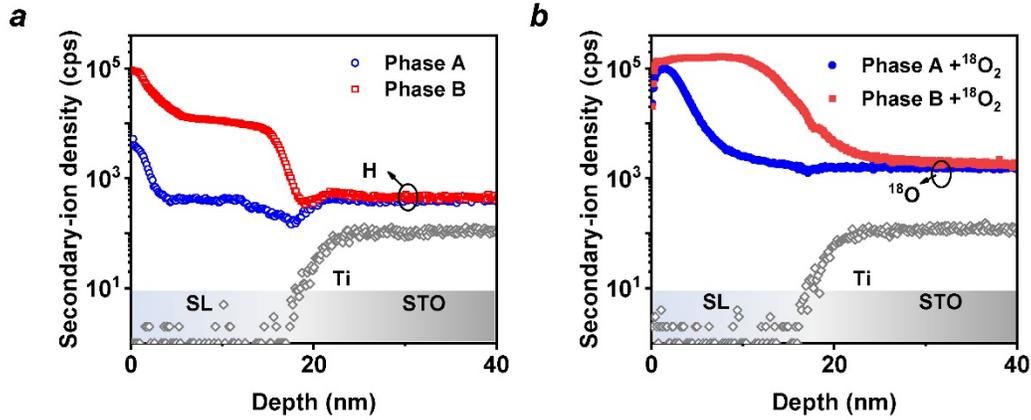

**Figure 2 | Chemical characterization.** (a) Depth profiles of H and Ti signals in the two phases of the superlattices, measured by using secondary-ion mass spectrometry. The H signal was obtained in phase A and B of the superlattices stabilized by ILG. The Ti signal from the substrate indicates the position of the interface between substrate and superlattice. (b) Depth profiles of $^{18}O$ and Ti signals in the two phases of the superlattices after thermal annealing in $^{18}O_2$. To measure $^{18}O$ signal, the superlattices were first stabilized in phase A and B under ILG, and then thermally annealed in $^{18}O_2$ gas.

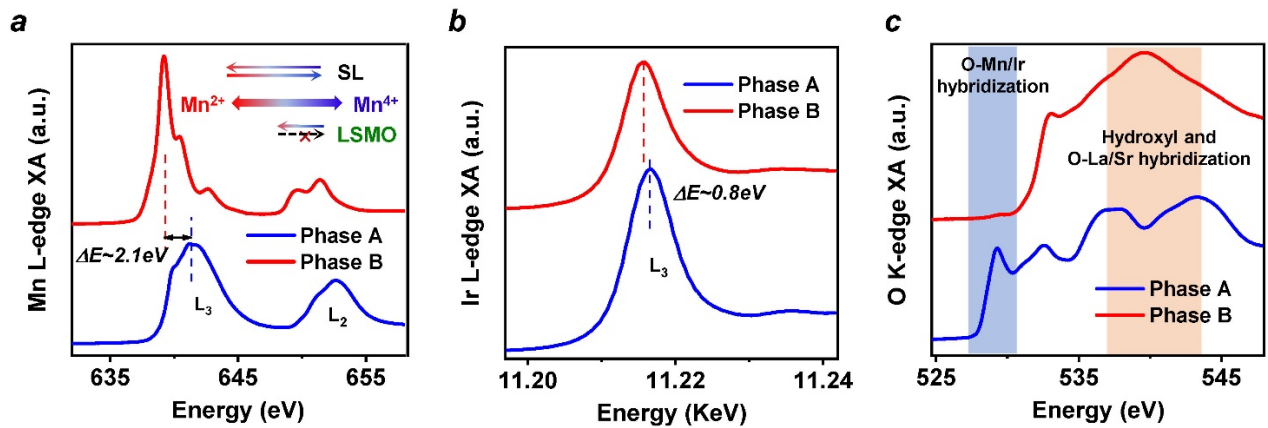

**Figure 3 | Valence characterization.** X-ray absorption (XA) spectra of (a) Mn L-edge, (b) Ir $L_3$-edge and (c) oxygen K-edge of the two phases. Figure 3a inset shows the valence changes of the superlattices and

La$_{0.2}$Sr$_{0.8}$MnO$_3$ films under ILG. The shifts of absorption peaks at the Mn L-edge (a) and Ir L$_3$-edge (b) reveal a significant decrease of valence for both Mn and Ir cations in phase B. The oxygen K-edge results (c) show a full suppression of hybridization between oxygen 2$p$ orbitals and Mn/Ir $d$ orbitals, and the appearance of features associated with hydroxyl bonds in phase B.

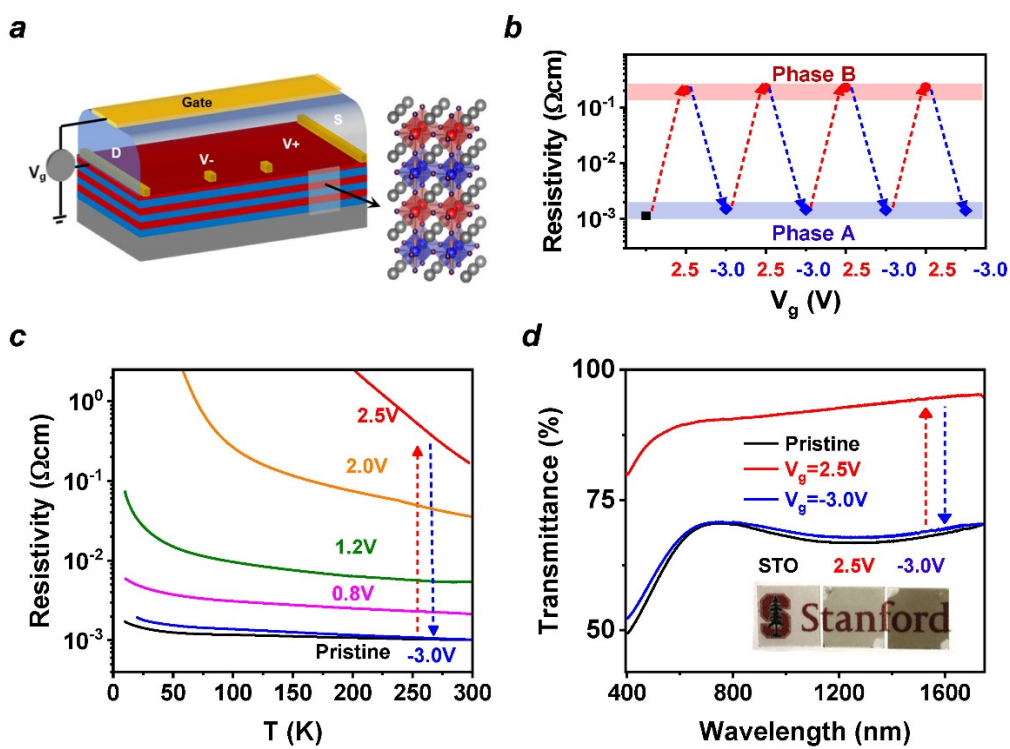

**Figure 4 | Electric field control of resistivity and electrochromic effect.** (a) Schematic of the device for in-situ measurements of the transport properties under ILG. (b) Reversible modulation of resistivity between the two phases during repeated voltage cycling at room temperature. (c) Temperature dependence of resistivity of the superlattice during voltage cycling (incremental positive voltages to +2.5 V and -3 V). (d) Optical transmittance spectra of a superlattice during voltage cycling. Inset shows the photographs of a double-side polished SrTiO$_3$ substrate and superlattices that are stabilized in the two phases by ILG.

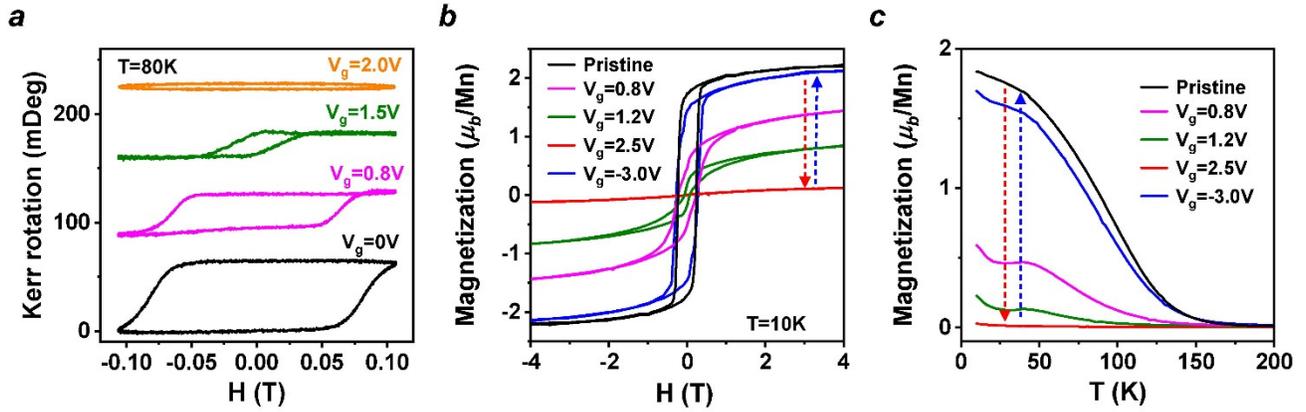

**Figure 5 | Electric field control of magnetic properties.** (a) Magnetic hysteresis loops of a superlattice under ILG, probed with in-situ magneto-optic Kerr effect (MOKE). An offset is applied for illustration. (b) Magnetic hysteresis loops and (c) temperature dependence of magnetization of the superlattice during voltage cycling, probed with ex-situ SQUID magnetometry. All measurements were performed along the out-of-plane direction, which is the magnetic easy axis due to interface perpendicular magnetic anisotropy. A field of 0.2 T was applied to obtain results in (c).

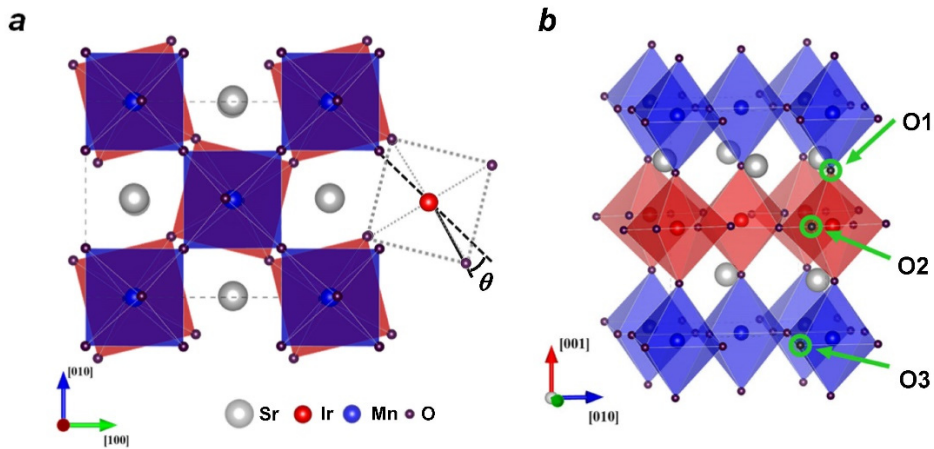

**Figure 6 | DFT calculations.** (a) Top view of the superlattice without ionic defects. The $IrO_6$ and $MnO_6$ octahedra of the superlattice show different magnitudes of rotation. The calculated rotation angle ($\theta$) along

the **c** axis ([001]) is about 14.5° and 2.9° for $IrO_6$ and $MnO_6$ octahedra, respectively. (b) Side view of the superlattice with the three kinds of oxygen vacancy sites. Due to the atomically layered structure, three distinct oxygen sites were considered, labelled as O1 (in the SrO layer), O2 (in the $IrO_2$ layer) and O3 (in the $MnO_2$ layer). Further results on hydrogen interstitial are included in the Supplementary Note 13.